# Physics-Informed Adaptive Deep Koopman Operator Modeling for Autonomous Vehicle Dynamics

Jianhua Zhang[a], Yansong He[a] and Hao Chen[a,b,*]

[a]*College of Mechanical and Vehicle Engineering, Chongqing University, Chongqing, 400044, China*
[b]*School of Mechanical and Aerospace Engineering, Nanyang Technological University, Singapore, 639798, Singapore*

ABSTRACT

Koopman operator has been recognized as an ongoing data-driven modeling method for vehicle dynamics which lifts the original state space into a high-dimensional linear state space. The deep neural networks (DNNs) are verified to be useful for the approximation of Koopman operator. To further improve the accuracy of Koopman operator approximation, this paper introduces a physical loss function term from the concept of physics-informed neural networks (PINNs), i.e., the acceleration loss between neural network output and sensor measurements, to improve the efficiency of network learning and its interpretability. Moreover, we utilize the sliding window least squares (SWLS) to update the system matrix and input matrix online in the lifted space, therefore enabling the deep Koopman operator to adapt to the rapid dynamics of autonomous vehicles in real events. The data collection and validation are conducted on CarSim/Simlink co-simulation platform. With comparison to other physics-based and data-driven approaches on various scenarios, the results reveal that the acceleration loss-informed network refines the accuracy of Koopman operator approximation and renders it with inherent generalization, and the SWLS enforces the deep Koopman operator's capability to cope with changes in vehicle parameters, road conditions, and rapid maneuvers. This indicates the proposed physics-informed adaptive deep Koopman operator is a performant and efficient data-driven modeling tool.

## 1. INTRODUCTION

Autonomous driving consists of five main components: perception, decision-making, control, communication, and vehicle dynamics modeling [1, 2, 3, 4, 5]. As for low-speed scenarios, classical kinematic models are commonly employed to capture vehicle motions [6]. However, in scenarios such as high-speed driving, tilting, drifting, etc., it is difficult to model the vehicle dynamics accurately due to the coupling between the lateral and longitudinal motions, as well as the strong nonlinearity of the tire characteristics. In light of the aforementioned reasons, researchers have proposed physics-based models, including the 7 DoF (degree-of-freedom) [7], the 14-DoF [8], and even the 17-DoF [9] to feature the behavior of vehicles in such scenarios. An increase in the degree of freedom of the model necessitates more precise knowledge in hyper-parameters (e.g., the center of gravity, the lateral cornering stiffness of tires, the road adhesion coefficient, etc.). This continues to be a challenging task in real events.

Data-driven methodology has exhibited satisfactory performance in modeling, optimization, and control. Researchers have incorporated machine learning techniques into vehicle dynamics systems [10, 11]. The driving force behind their work is the utilization of raw data, which is employed to train network models in order to construct vehicle models that are approximations. For instance, Spielberg [12] adopted a neural network and trained it by collections data on dry and icy roads to predict the road adhesion coefficient, which simplified the estimation of road adhesion coefficient. Nevertheless, the neural network's interpretability and generalization are insufficient because of its "black box" nature. Besides, the developed model lacks compatibility with further controller design. Among this, Koopman operator, named after Bernard Koopman, has received significant attention in recent years [13], where the nonlinear system in the original space is augmented into a linear form in the lifted space through an infinite-dimensional operator acting on system measurements. This enables the possibility of applying analysis and synthesis techniques for linear systems to realize modeling and control of nonlinear systems. The infinite-dimensional operator is impossible to deploy in real applications.

*Corresponding author
✉ 202207131490@stu.cqu.edu.cn (J. Zhang); hys68@cqu.edu.cn (Y. He); h-c15@outlook.com (H. Chen)
ORCID(s): 0000-0003-4649-5507 (H. Chen)

Rowley et al. [14] first introduced Dynamic Mode Decomposition (DMD) to approximate the Koopman operator by singular value decomposition (SVD). The approach was initially employed by Schmid within fluid mechanics and has been a standard algorithm for neuroscience [15], robotics [16], and power systems [17]. However, the DMD fails to feature nonlinearities because of its inherent linear transformations. Williams et al. [18] proposed the Extended Dynamic Mode Decomposition (EDMD) technique as a solution. The EDMD primarily applies linear DMD logic to the extended nonlinear basis function, which has been proven to be more effective in Koopman operator approximation. Cibulka et al. [19] investigated the EDMD to capture vehicle dynamics, resulting in successful outcomes. Svec et al. [20] developed a model predictive controller identified by the Koopman operator through the EDMD approach for obtaining the optimal input in the form of longitudinal and lateral tire forces. Compared to linearization by Taylor expansion, the Koopman operator-based model has excellent accuracy for vehicle dynamics. Besides, the selection of nonlinear basis functions of EDMD depends on the expertise of researchers or the case-by-case basis, which impacts the model fidelity of Koopman operator approximation and lacks generalization ability[21].

Machine Learning (ML) efficiently represents nonlinear functions and reveals promising potential in Koopman operator approximation [22]. Many scholars incorporated ML with the Koopman operator to simulate and regulate vehicle behaviors. Wang et al. [23] adopted the deep neural network (DNN) to construct the basis function of Koopman operator. The trained deep Koopman operator model was then used as a predictive model to follow the desired trajectory. The proposed controller presented superior tracking performance than the nonlinear kinematic-based model. Besides, the deep Koopman model also exhibits good robustness. Xiao et al. [24] employed DNN-based Koopman operator to model the dynamics of autonomous vehicle. The encoder and decoder nets were built during the training and validation, thereby mitigating the limitations of the manual selection of basis functions. The effectiveness of this strategy in terms of tracking accuracy and computation efficiency has been validated by experimental investigations. On the other hand, DNN is insufficient to extract interpretable information from extensive datasets despite its remarkable performance in a variety of contexts. In addition, autonomous vehicles operate in various scenarios with changing mass and moment of inertial, characterized as typical time-varying systems. Generalization issues may occur when the feature is not present in the training dataset, resulting in unexpected predictions. In recent times, there has been a growing awareness of these points, raising the development of a new machine learning algorithm called Physics-Informed Neural Networks (PINNs) [25]. PINNs makes full use of prior physical knowledge or constraints within the dataset to generate more interpretable results, which improves the robustness of neural networks and aligns with general physical principles in their predictions. Misyris et al. [26] applied PINNs to the power system for describing power system behavior. The computational efficiency of rotor angle and frequency is 87 times than that of the traditional method with less data. Xu et al. [27] used PINNs to identify the inertia and damping coefficient of unmanned surface vehicles (USV) instead of DNN. The speed (motion) and steering dynamics of the USV are embedded into the loss function to respect physical constraints and improve efficiency of neural network learning. The experimental results indicate that the PINNs have better generalization ability to predict the surge, sway velocities and rotation speed than than other incumbent ML algorithms.

Furthermore, least squares-based methods such as recursive least squares (RLS) [28], forgetting factor recursive least squares (FFRLS) [29], and sliding window least squares (SWLS) [30], are able to estimate time-varying parameters in real-time. Towliat et al. [31] introduced a multi-layer RLS estimator for fast time-varying systems. By eliminating erroes from previous layer, the estimation accuracy is improved. Besides, this method also offers a moderate computational complexity. Wang et al. [32] proposed FFRLS to addresses data fluctuations caused by various factors during battery operation. It is verified that the FFRLS has more accurate estimations and stable performance on parameter identification. Retianza et al. [33] presented a SWLS method to compensate current sensor errors and mitigate computational load, while ensuring accuracy in highly dynamic three-phase electric drives.

Motivated by these efforts, this study introduces a physical loss function term from the concept of PINNs during Koopman operator approximation, i.e., the acceleration loss between neural network output and sensor measurement, to extract the interpretable information of the dataset. Moreover, to improve the adaptive capability of the autonomous vehicle model based on the deep Koopman operator, the SWLS algorithm is employed to online update system matrices in the lift space with comparisons to the RLS and FFRLS. Therefore, the main contributions of this paper can be concluded as:

(1) Introducing PINNs into the deep Koopman operator framework, and adding a physics-informed loss, i.e., acceleration loss, to improve the efficiency of network learning and its interpretability. This enhances the accuracy of deep Koopman operator approximation and modeling accuracy of autonomous vehicle dynamics.

(2) Using SWLS to update the system matrices in the lifted space augmented by the deep Koopman operator in a real-time manner; selecting an appropriate window size that can balance the impact of outdated historical and current data. This enforces the deep Koopman operator model adaptive to the time-varying dynamics of autonomous vehicles.

The remainder of this paper is organized as follows: Section 2 gives an overview of the concept of Koopman operator, deep Koopman operator, and SWLS. Section 3 elaborates on the physics-informed deep Koopman model for autonomous vehicle dynamics, including the proposed acceleration loss added into network training and the online update system matrices in the lifted space by SWLS. The simulation tests and comparative studies are conducted and discussed on the various scenarios in Section 4, and conclusions are summarized in Section 5.

## 2. PRELIMINARIES

This section will primarily provide an introduction to the pertinent concepts of the Koopman operator, Deep-Koopman, and SWLS algorithm, additionally, we will provide several symbols that will be applied in the following chapters for the purpose of derivation.

### 2.1. Koopman Operator Theory

A general discrete-time nonlinear system is:

$$\boldsymbol{x}_{k+1} = \boldsymbol{f}(\boldsymbol{x}_k, \boldsymbol{u}_k) \tag{1}$$

where $\boldsymbol{x}_k \in \mathbb{R}^n$ is the state vector at time $k$, $\boldsymbol{u}_k \in \mathbb{R}^m$ is the control vector at time $k$, and $\mathbb{R}^{n+m} \to \mathbb{R}^n$ is a smooth vector field of $\boldsymbol{C}^\infty$.

Koopman operator, $\mathcal{K}$, is an infinite linear operator, spanned by Koopman observation functions or embedding functions $\boldsymbol{\varphi}$ in the lifted space as:

$$\mathcal{K} \cdot \boldsymbol{g}(\boldsymbol{x}_k, \boldsymbol{u}_k) = \boldsymbol{g}(\boldsymbol{x}_{k+1}, \boldsymbol{u}_{k+1}) = \boldsymbol{g}(\boldsymbol{f}(\boldsymbol{x}_k, \boldsymbol{u}_k), \boldsymbol{u}_{k+1}) \tag{2}$$

where $\boldsymbol{g}: \mathbb{R}^{n+m} \to \mathbb{R}^d$ is a smooth vector field of $\boldsymbol{C}^\infty$.

The embedding function $\boldsymbol{g}(\cdot)$ can be further separated into two parts in terms of states and control inputs.

$$\boldsymbol{g}(\boldsymbol{x}_k, \boldsymbol{u}_k) = \begin{bmatrix} \boldsymbol{g}_x(\boldsymbol{x}_k) \\ \boldsymbol{u}_k \end{bmatrix} \tag{3}$$

where $\boldsymbol{g}_x : \mathbb{R}^n \to \mathbb{R}^{d-m}$.

However, the implementation of the infinite-dimensional $\mathcal{K}$ is impractical, but it can be realized through a finite-dimensional approximation [34]. Thus, (2) is simplified without loss of generality as [35]:

$$\begin{bmatrix} \boldsymbol{g}_x(\boldsymbol{x}_{k+1}) \\ \boldsymbol{u}_{k+1} \end{bmatrix} = \begin{bmatrix} \boldsymbol{K}_{xx} & \boldsymbol{K}_{xu} \\ \boldsymbol{K}_{ux} & \boldsymbol{K}_{uu} \end{bmatrix} \begin{bmatrix} \boldsymbol{g}_x(\boldsymbol{x}_k) \\ \boldsymbol{u}_k \end{bmatrix} \tag{4}$$

where $\boldsymbol{K} \in \mathbb{R}^{d \times d}$ is the finite-dimension approximation of $\mathcal{K}$ in the lifted space.

From (4), we denoted, and the state evolution in the lifted space can be written as:

$$\boldsymbol{z}_{k+1} = \boldsymbol{A} \boldsymbol{z}_k + \boldsymbol{B} \boldsymbol{u}_k \tag{5}$$

where $\boldsymbol{A} = \boldsymbol{K}_{xx} \in \mathbb{R}^{(d-m) \times (d-m)}$ and $\boldsymbol{B} = \boldsymbol{K}_{xu} \in \mathbb{R}^{(d-m) \times m}$ are the system and input matrices in the lifted space.

The component $\boldsymbol{z}_k = \boldsymbol{g}_x$ of the embedding function $\boldsymbol{g}$ is defined as:

$$\boldsymbol{z}_k = \boldsymbol{g}_x(\boldsymbol{x}_k) = \begin{bmatrix} \boldsymbol{x}_k \\ \boldsymbol{\Phi}(\boldsymbol{x}_k) \end{bmatrix} \tag{6}$$

where $\boldsymbol{\Phi}(\boldsymbol{x}_k) = [\varphi_{1,k}(\boldsymbol{x}_k), \varphi_{2,k}(\boldsymbol{x}_k), ..., \varphi_{d-m-n,k}(\boldsymbol{x}_k)]^T \in \mathbb{R}^{d-m-n}$ is the basis in the lifted space. Because the original states are included in the lifted space in (6), the original state space can be reconstructed without loss any accuracy by the $\boldsymbol{z}_k$ projection as:

$$\boldsymbol{x}_k = \boldsymbol{C} \boldsymbol{z}_k \tag{7}$$

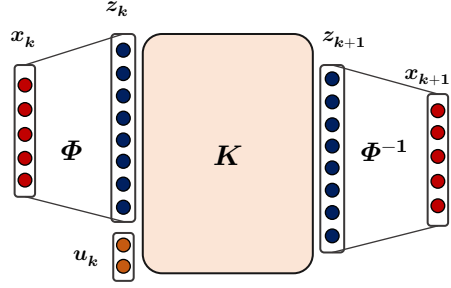

**Figure 1:** DNN-based Koopman operator (deep Koopman operator) approximation.

where $C = \begin{bmatrix} I^n & 0^{n\times(d-m-n)} \end{bmatrix}$ is the projection matrix from the lifted space to the original space, $I^n$ is an $n$-dimensional unit matrix.

Further, the $l$-snapshots collections from the lifted space are:

$$\begin{cases} Z_X = [z_1, z_2, \cdots, z_k]_{(d-m)\times l} \\ U = [u_1, u_2, \cdots, u_l]_{m\times l} \\ Z_Y = [z_2, z_3, \cdots, z_{l+1}]_{(d-m)\times l} \end{cases} \qquad (8)$$

where $l$ is the sequence length. The objective function is defined as:

$$\min_{A,B} \|Z_Y - AZ_X - BU\|_2^2 \qquad (9)$$

The formal simplified expression as:

$$\begin{bmatrix} A & B \end{bmatrix} = Z_Y \begin{bmatrix} Z_X \\ U \end{bmatrix}^T \left( \begin{bmatrix} Z_X \\ U \end{bmatrix} \begin{bmatrix} Z_X \\ U \end{bmatrix}^T \right)^\dagger \qquad (10)$$

where † denotes pseudo-inverse.

### 2.2. DNN for Koopman Operator Approximation

Although it is not difficult to solve the above linear least squares problem in (10), the selection of basis function $\boldsymbol{\Phi}$ is still a challenging task. Eq.(6) indicates that $\boldsymbol{\Phi}$ has an influence on the linearity in the lifted space and later modeling accuracy. As stated in Section 1, DNN provides a tutorial on the definition of basis function $\boldsymbol{\Phi}$.

Within the framework of DNN-based Koopman operator (deep Koopman operator) in Figure 1, a lightweight auto-encoder and auto-decoder network structure, rather than manual selection, is adopted for the basis function $\boldsymbol{\Phi}$, due to its nonlinearity and unknowns. The original space is lifted into the high-dimensional state space via $g_x$, which evolves in linear dynamics. The encoder or decoder is composed of multi-layer perceptron [36] approximation $\boldsymbol{\Phi}$ or $\boldsymbol{\Phi}^{-1}$, with a general form of the $j$-th hidden layers as:

$$y_j = \sigma_j(W_j y_{j-1} + b_j) \quad (j = 1, ..., J) \qquad (11)$$

where $W_j \in \mathbb{R}^{n_j \times n_{j-1}}$ and $b_j \in \mathbb{R}^{n_j}$ is the weight matrix and bias vector of the $j$-th hidden layer, respectively, $\sigma_j$ is the activation function of the $j$-th hidden layer, and $n_j$ is the number of neurons in the $j$-th hidden layer.

Besides, two single-layer networks without bias are constructed to identify system matrix $A$ and input matrix $B$ in the lifted space, respectively. The encoder $\boldsymbol{\Phi}$, decoder $\boldsymbol{\Phi}^{-1}$, system matrix $A$ and input matrix $B$ are updated synchronously during the training process. The loss function often consists of three terms:

(1) **Linearity loss:** linearity of the lifted space,

$$L_{\text{Linear}} = \left\| \begin{bmatrix} x_{i+1} \\ \Phi(x_{i+1}) \end{bmatrix} - \begin{bmatrix} A & B \end{bmatrix} \begin{bmatrix} x_i \\ \Phi(x_i) \\ u_i \end{bmatrix} \right\| \qquad (12)$$

(2) **Reconstruction loss:** the consistency after the reconstruction of encoder and decoder,

$$L_{\text{Recon}} = \|x_i - \Phi^{-1}(\Phi(x_i))\| \tag{13}$$

(3) **Prediction loss:** the accuracy of the prediction in the original space after evolution in the lifted space,

$$L_{\text{Pred}} = \left\| x_{i+1} - C \begin{bmatrix} A & B \end{bmatrix} \begin{bmatrix} \begin{bmatrix} x_i \\ \Phi(x_i) \end{bmatrix} \\ u_i \end{bmatrix} \right\| \tag{14}$$

**Remark 1.** Based on the loss function of (12)(13)(14), the DNN-based Koopman operator or deep Koopman operator is approximated after neural networks learning. as we addressed above, the deep Koopman operator avoids manual selection of basis function and enhance the linearity in the lifted space.

### 2.3. Sliding Window Least Squares (SWLS) Algorithm

Least squares algorithm and its variants have been widely used in system identification of autonomous vehicles [37, 38, 39]. Its recursive and batch form has satisfactory computational efficiency for online implementation. Compared to RLS and FFRLS, SWLS can drop historical data outside of a specific time window to better capture the time-varying dynamics of the system.

Specifically, for a general discrete-time linear system:

$$y = hx \tag{15}$$

where $y \in \mathbb{R}^{m \times 1}$ is the output vector, $x \in \mathbb{R}^{n \times 1}$ is the input vector, and $h \in \mathbb{R}^{m \times n}$ is the linear transformation matrix to be estimated.

The $l$-snapshots collection of (15) is:

$$Y = hX \tag{16}$$

where $Y \in \mathbb{R}^{m \times l}$ and $X \in \mathbb{R}^{n \times l}$ are denoted as:

$$\begin{aligned} Y &= \begin{bmatrix} y_{1,1} & \cdots & y_{1,l} \\ \vdots & \ddots & \vdots \\ y_{m,1} & \cdots & y_{m,l} \end{bmatrix} \\ X &= \begin{bmatrix} x_{1,1} & \cdots & x_{1,l} \\ \vdots & \ddots & \vdots \\ x_{n,1} & \cdots & x_{n,l} \end{bmatrix} \end{aligned} \tag{17}$$

When the number of snapshots is greater than the dimension of input, i.e., $l \gg n$, is $X$ full roe rank, and the least square solution of (16) is:

$$h = YX^T[XX^T]^{-1} \tag{18}$$

As for SWLS, a time window is formed over snapshot pairs. Meanwhile, this window slides over the data to capture the latest portions. The window length is set as $M$ to capture sequential snapshots over the last $M$ steps at time $k$ as:

$$\begin{aligned} Y_k &= \begin{bmatrix} y_{1,k-M+1} & \cdots & y_{1,k} \\ \vdots & \ddots & \vdots \\ y_{m,k-M+1} & \cdots & y_{m,k} \end{bmatrix} \\ X_k &= \begin{bmatrix} x_{1,k-M+1} & \cdots & x_{1,k} \\ \vdots & \ddots & \vdots \\ x_{n,k-M+1} & \cdots & x_{n,k} \end{bmatrix} \end{aligned} \tag{19}$$

Denotes $P_k = \left[ X_k X_k^T \right]^{-1}$, for simplicity, (18) is re-written as:

$$h_k = [Y_{k-1} \quad y_k][X_{k-1} \quad x_k]^T P_k \tag{20}$$

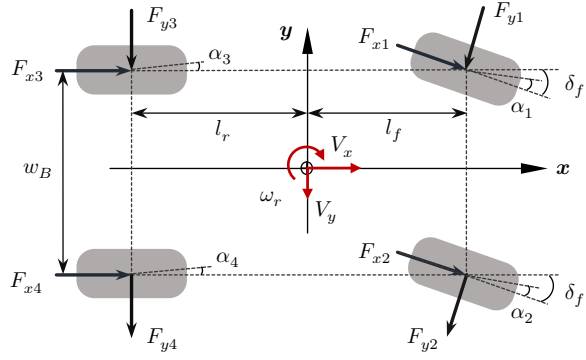

**Figure 2:** The planar dynamics model of a four wheel-driven vehicle.

Further, the recursive form of solution within the $M$-steps time window is:

$$h_k = h_{k-1} + [y_k - h_{k-1}x_k]x_k^T P_k \tag{21}$$

**Remark 2.** When the number of snapshots is less than the window length, $k < M$, $h_k$ is solved by (18). otherwise, $h_k$ is recursively updated as (21) with the initial value at time $M$, i.e., $h_M$. On the other hand, it is notable that, when $M=1$, the SWLS is equivalent to a general least squares algorithm, and if $M = k$, it becomes a general RLS.

## 3. PHYSICS-INFORMED ADAPTIVE DEEP KOOPMAN OPERATOR for VEHICLE DYNAMICS MODELING

### 3.1. Vehicle Dynamics Model

A four wheel-driven vehicle [40] is taken as the objection in Figure 2. The Newton-Euler equations of planner motion for the vehicle are:

$$\begin{cases} \dot{V}_x = \dfrac{1}{m}\left[(F_{x1}+F_{x2})\cos\delta_f - (F_{y1}+F_{y2})\sin\delta_f\right] + \dfrac{1}{m}(F_{x3}+F_{x4}) + V_y\omega_r \\ \dot{V}_y = \dfrac{1}{m}\left[(F_{x1}+F_{x2})\sin\delta_f - (F_{y1}+F_{y2})\cos\delta_f\right] + \dfrac{1}{m}(F_{y3}+F_{y4}) - V_x\omega_r \\ \dot{\omega}_r = \dfrac{w_B}{2I_z}\left[(F_{x2}\cos\delta_f - F_{y2}\sin\delta_f) + F_{x4} - (F_{x1}\cos\delta_f - F_{y1}\sin\delta_f) - F_{x3}\right] \\ \quad + (F_{x2}\sin\delta_f + F_{y2}\cos\delta_f + F_{x1}\sin\delta_f + F_{y1}\cos\delta_f)\dfrac{l_f}{I_z} - (F_{y3}+F_{y4})\dfrac{l_r}{I_z} \end{cases} \tag{22}$$

where $V_x$, $V_y$, and $w_r$ are longitudinal velocity, lateral velocity, and yaw rate of the vehicle, respectively. $F_{xi}$ ($i = 1, 2, 3, 4$) is the longitudinal tire force, $F_{yi}$ ($i = 1, 2, 3, 4$) is the lateral tire force, $\delta_f$ is the steering angle of the front wheel, $m$ is the vehicle mass, $I_z$ is the mass moment, $l_f$ ($l_r$) is the distance from the front (rear) axle to c.g., and $w_B$ is the track width.

The state and input vectors are defined as:

$$\begin{aligned} x &= \begin{bmatrix} V_x & V_y & \omega_r \end{bmatrix}^T \\ u &= \begin{bmatrix} T & \delta_f \end{bmatrix}^T \end{aligned} \tag{23}$$

where $T$ is the driving torque, that is, the summation of motor torque or braking torque of each wheel. The longitudinal tire forces $F_{xi}$ ($i = 1, 2, 3, 4$) and lateral tire forces $F_{yi}$ ($i = 1, 2, 3, 4$) are the implicit functions with respect to $x$ and $u$ [41, 42, 43].

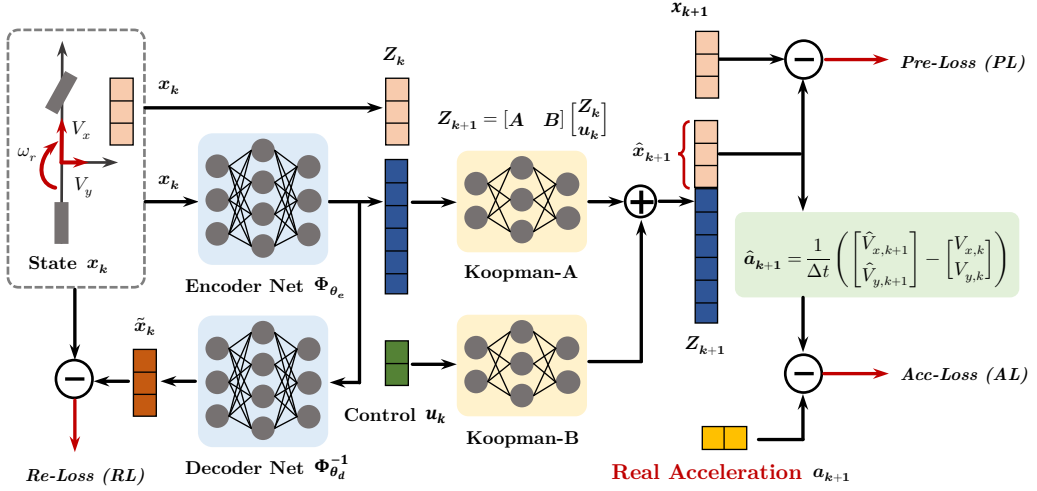

**Figure 3:** Physics-informed deep Koopman operator approximation for vehicle planar dynamics.

As such, based on (23), the discrete-time form of (22) is denoted as:

$$\begin{bmatrix} V_{x,k+1} \\ V_{y,k+1} \\ \omega_{r,k+1} \end{bmatrix} = f\left( \begin{bmatrix} V_{x,k} \\ V_{y,k} \\ \omega_{r,k} \end{bmatrix}, \begin{bmatrix} T_k \\ \delta_{f,k} \end{bmatrix} \right) \tag{24}$$

where $f(\cdot)$ maps the nonlinear evolution dynamics of the vehicle in a planar motion consistent with that of (1).

### 3.2. Physics-informed Adaptive Deep Koopman Operator Model

PINNs leverage the knowledge of physical laws to extrapolate solutions beyond the data points, which represent a promising approach for solving complex, physics-based problems efficiently and effectively. The framework of physics-based deep Koopman operator approximation for vehicle planer dynamics is introduced in Figure 3. As we stated in Section 2.2, the auto-encoder and auto-decoder network structures are used to learn the basis function of Koopman operator. The decoder has a similar structure as the encoder, which forces the encoder net to capture the main facets of the input data so that the decoder can reconstruct the input based on the encoder features.

#### 3.2.1. Physics-informed Loss - Acceleration Loss

Autonomous vehicles are often equipped with GPS, inertial measurement unit (IMU), and other high-resolution sensors that can provide longitudinal and lateral acceleration signals in the body frame. By utilizing the current states and the one-step further prediction states derived from the projection of the lifted space, it is possible to calculate the changing rate of longitudinal and lateral velocity. In contrast, the physical measurement of these states can be directly obtained. We defined the new physical-informed loss term, i.e., acceleration loss, as:

$$L_{\text{AL}} = \|a_{k+1} - \hat{a}_{k+1}\| \tag{25}$$

where $\hat{a}_{k+1}$ is the changing rate of velocity based on the prediction of Koopman model as follows:

$$\hat{a}_{k+1} = \frac{1}{\Delta t}\left( \begin{bmatrix} \hat{V}_{x,k+1} \\ \hat{V}_{y,k+1} \end{bmatrix} - \begin{bmatrix} V_{x,k} \\ V_{y,k} \end{bmatrix} \right) \tag{26}$$

and,

$$a_{k+1} = \begin{bmatrix} a_{x,k+1} + V_{y,k+1}\omega_{r,k+1} \\ a_{y,k+1} - V_{x,k+1}\omega_{r,k+1} \end{bmatrix} \tag{27}$$

where $a_{x,k+1}$ and $a_{y,k+1}$ is the longitudinal and lateral acceleration in the body frame from sensors at time $k+1$, respectively.

Since the original state space has been embedded into the lifted space, $\hat{V}_{x,k+1}$ and $\hat{V}_{y,k+1}$ are the first two elements of the one-step prediction state vector $z_{k+1}$. Therefore, incorporating the new term (25) with (12), (13), (14), the total loss is defined as:

$$L_{\text{Total}} = L_{\text{Linear}} + L_{\text{Recon}} + L_{\text{Pred}} + L_{\text{AL}} \tag{28}$$

The total loss function renders the network not only map nonlinear vehicle dynamics but also compliance with underlying physical principles, thus improving its accuracy and interpretability.

After undergoing sufficient training, the network parameters $\theta_e$, $\theta_d$ of encoder and decoder, as well as the system matrix $\boldsymbol{A}$ and the input matrix $\boldsymbol{B}$, have been identified.

$$\min_{\theta_e,\theta_d,\boldsymbol{A},\boldsymbol{B}} L_{\text{Total}} \tag{29}$$

### 3.2.2. Adaptive Update of Deep Koopman Operator with SWLS

The SWLS is a valuable method that balances memory usage with adaptation to the time-varying system. As discussed in Section 2.3, when the current time-step $k$ is less than the window length, all the data are recorded, and then the SWLS is equivalent to the general RLS. Conversely, when the current time-step exceeds the window length, the $M$-snapshots pairs are saved and progressively updated. Finally, through Equation , the approximate matrix can be updated recursively. The details are as follows:

(1) $k < M$ (the number of snapshots is less than the window length)

$$\begin{cases} \boldsymbol{\psi} = [\boldsymbol{z}_0,\cdots,\boldsymbol{z}_{k-1}]_{(d-m)\times k} \\ \boldsymbol{U} = [\boldsymbol{u}_0,\cdots,\boldsymbol{u}_{k-1}]_{m\times k} \end{cases} \tag{30}$$

(2) $k \geq M$ (the number of snapshots is equal or more than the window length)

$$\begin{cases} \boldsymbol{\psi} = [\boldsymbol{z}_{k-M},\cdots,\boldsymbol{z}_{k-1}]_{(d-m)\times M} \\ \boldsymbol{U} = [\boldsymbol{u}_{k-M},\cdots,\boldsymbol{u}_{k-1}]_{m\times M} \end{cases} \tag{31}$$

Based on the (19), $\boldsymbol{P}_{k-1}$ and its initial value $\boldsymbol{P}_0$ are defined as:

$$\begin{cases} \boldsymbol{P}_{k-1} = \left[\begin{bmatrix}\boldsymbol{\psi}\\\boldsymbol{U}\end{bmatrix}[\boldsymbol{\psi}^T\ \boldsymbol{U}^T]\right]^{-1} & (k \geq 1) \\ \boldsymbol{P}_0 = \left[\begin{bmatrix}\boldsymbol{\psi}_0\\\boldsymbol{U}_0\end{bmatrix}[\boldsymbol{\psi}_0^T\ \boldsymbol{U}_0^T]\right]^{-1} & (k=0) \end{cases} \tag{32}$$

As a result, the system and input matrices in the lifted space can be updated recursively as:

$$[\boldsymbol{A}_k\ \ \boldsymbol{B}_k] = [\boldsymbol{A}_{k-1}\ \ \boldsymbol{B}_{k-1}] + \left[\boldsymbol{z}_k - [\boldsymbol{A}_{k-1}\ \ \boldsymbol{B}_{k-1}]\begin{bmatrix}\boldsymbol{z}_{k-1}\\\boldsymbol{u}_{k-1}\end{bmatrix}\right]\boldsymbol{P}_{k-1} \tag{33}$$

Notably, the initial value of $\boldsymbol{A}_0$ and $\boldsymbol{B}_0$ are the results of (29) after training. The adaptive update of the deep Koopman operator with SWLS is summarized in Algorithm 1.

**Remark 3.** All vehicle-related parameters (such as tire characteristics, center of height, etc.) are not predetermined in this article. The vehicle dynamics model is constructed solely based on the deep Koopman data-driven method. Vehicle states and control input data are used to mine the potential mapping relationship of the vehicle dynamics system under the representation of (1) using the neural network capability to establish a model in place of the conventional physical model by collection under specific working conditions.

**Algorithm 1** Adaptive Update of Deep Koopman Operator with SWLS

1: **Initialization:**, $A_0$, $B_0$, $z_0$, $u_0$, $P_0$, Window Length $M$.
2: **for** $k=1$:end **do**
3:    Collect state $x_k$ and control input $u_k$ at time $k$.
4:    Calculate $z_k$ with (6).
5:    **if** $k < M$ **then**
6:      Record state matrix $\psi$ and control input matrix $U$ with (30).
7:    **else**
8:      Record state matrix $\psi$ and control input matrix $U$ with (31).
9:    **end if**
10:   Calculate $P_k$ with (32).
11:   Update the approximation of $A_k$ and $B_k$ with (33).
12: **end for**

**Table 1**
The configurations of IMDV in CarSim

| Parameter | Nomenclature | Value | Uint |
|---|---|---|---|
| Distance from c.g. to Front axle | $l_f$ | 1.315 | m |
| Distance from c.g. to Rear axle | $l_r$ | 1.355 | m |
| Track Width | $w_B$ | 1.715 | m |
| Vehicle Mass | $m$ | 2070 | kg |
| Yaw Inertia | $I_z$ | 3658 | $kg \cdot m^2$ |

## 4. VALIDATION AND DISCUSSION

In this section, we choose In-wheel Motor Driven Vehicle (IMDV) as the control plant and feature the IMDV's partial planar dynamics using the data-driven model—physics-informed adaptive deep Koopman operator. Further, the fidelity, robustness, and generalization of proposed deep Koopman operator are validated by comparison with benchmarks from CarSim/Simulink platform.

### 4.1. Data Collection, Preprocess and Model Training

The configurations of IMDV are listed in Table 1. The *X-Y-Z edges* in CarSim defines the horizontal geometry and elevation information of the reference path. which features the large and small curvature steering conditions, straight road and a vertically undulating road surface, the road adhesion coefficient is 0.85. This allows for a more accurate representation of real-world events. Therefore, we conduct the data collection on the *X-Y-Z edges* road. Besides, no extra restrictions are imposed on control inputs other than following the reference path. With a simulation duration of 1400s and a sampling interval of 0.025s, 56,001 snapshots are collected (where 70% are used as the training dataset and the remaining 30% as the testing dataset). Each snapshot consists of five states ($V_x$, $V_y$, $\omega_r$, $a_x$, $a_y$), and two control inputs ($T$, $\delta_f$).

During the training process, the auto-encoder and auto-decoder networks comprise three hidden layers with 128 neurons. Two single-layer network without bias in the lifted space as the system matrix $A$ and input matrix $B$ are also built therein. In addition, we apply the maximum-minimum normalization method [44] to ensure features are within a similar scale and mitigate vanishing or exploding gradients. The structure and data scope of the networks are detailed in Table 2.

### 4.2. Model Fidelity Validation

We first verify the accuracy of the acceleration loss-informed deep Koopman operator model (ALDK) compared to the deep Koopman operator model (DK) without this physics loss. Then, we underline the adaption capability of our proposed Koopman operator in interaction with fast-varying vehicle dynamics over other methods.

**Table 2**
The training hype-parameters of neural networks

| Hyper-Parameter | Value |
| --- | --- |
| Learning Rate | $10^{-3}$ |
| Batch Size | 20 |
| Epoch | 2000 |
| Auto-encoder | $\begin{bmatrix} 3 & 128 & 128 & 128 & 12 \end{bmatrix}$ |
| Auto-decoder | $\begin{bmatrix} 12 & 128 & 128 & 128 & 3 \end{bmatrix}$ |
| System matrix $A$ | $\begin{bmatrix} 15 & 15 \end{bmatrix}$ |
| Input matrix $B$ | $\begin{bmatrix} 2 & 15 \end{bmatrix}$ |

**Table 3**
Statistic results of deep-Koopman model with acceleration loss

| Estimation Error | $V_x$ | $V_y$ | $\omega_r$ |
| --- | --- | --- | --- |
| | Max/RMSE | Max/RMSE | Max/RMSE |
| DK | 2.030/0.620 | 0.228/0.090 | 4.844/1.552 |
| **ALDK** | **1.840/0.220** | **0.053/0.014** | **1.157/0.282** |

### 4.2.1. Acceleration loss-informed Deep Koopman Operator

Both the ALDK and DK show the overall trend with the benchmark in Figure 4. To further elaborate on the improvement of ALDK, we examine the estimation errors of longitudinal velocity $V_x$, lateral velocity $V_y$, and yaw rate $\omega_r$ in Table 3, respectively.

The ALDK has notable enhancements in velocities and yaw rate of the IMDV over the DK model. Compared to the DK, the maximum estimation error of yaw rate for the ALDK decreases from 4.844 deg/s to 1.157 deg/s, and its root mean square error (RMSE) exhibits an 87% reduction. As for the longitudinal velocity, the ALDK presents a maximum estimation error of 1.840 km/h and a RMSE of 0.220 km/h, whereas that of the DK are 2.030 km/h and 0.620 km/h, respectively. Similarly, the RMSE of lateral velocity of the ALDK has also been lowered by 89%.

Overall, the deep Koopman operator incorporating acceleration loss has a more satisfactory performance for the modeling of the IMDV's planar dynamics. Using the acceleration term as a physical loss when training neural networks can enhance the accuracy of deep Koopman operator approximation and modelling vehicle behaviors.

### 4.2.2. Online Adaptability

In this section, five different approaches are evaluated for adaptability to the fast-varying dynamics of the IMDV and external environment:

(1) **7 DoF-MF:** a nonlinear physical model to depict the vehicle planar dynamics, proposed in [45]. The magic formula is adopted to obtain the lateral tire forces as the function of tire slip angle and road friction coefficient.
(2) **ALDK:** based on the learned acceleration loss-informed deep Koopman operator, the system matrix $A$ and input matrix $B$ in the lifted space are constant.
(3) **ALDK-RLS:** the system matrix $A$ and input matrix $B$ are updated recursively using the RLS
(4) **ALDK-FFRLS:** similar to 3), but the FFRLS is introduced therein.
(5) **ALDK-SWLS:** as outlined in Section 3, the SWLS is implemented to online reconfigure the system matrix $A$ and input matrix $B$.

It is notable that the testing dataset is the same as Section 4.2.1. The validation results are demonstrated in Figure 5 and Table 4.

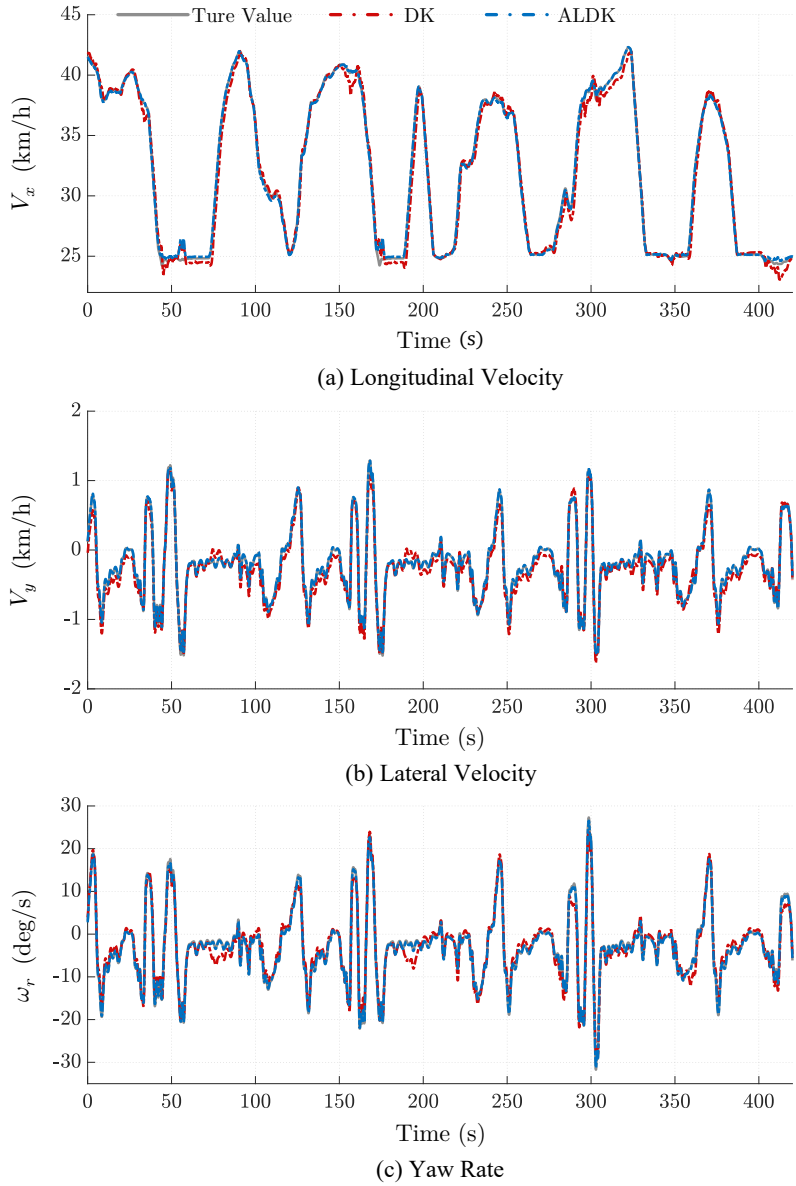

**Figure 4**: Simulation results of deep-Koopman model with acceleration loss.

The RMSE of longitudinal velocity, lateral velocity, and yaw rate for the ALDK are 0.220 km/h, 0.014 km/h, and 0.282 deg/s, respectively. The ALDK-RLS does not have a notable improvement over the ALDK. In contrast, the ALDK-FFRLS and ALDK-SWLS both provide enhancements in modeling accuracy of the IMDV planar dynamics. The 7 DoF-MF presents a satisfactory estimation for longitudinal velocity, but the results for lateral velocity and yaw rate are the worst among the five approaches. More specifically, the ALDK-SWLS algorithm demonstrates a substantial reduction in RMSE of 95%, 91%, and 91% than the ALDK for longitudinal velocity, lateral velocity, and yaw rate, respectively.

On the other hand, the maximum estimation error of the ALDK-RLS for yaw rate decreases from 1.157 deg/s of the ALDK to 0.960 deg/s, but its estimation in the longitudinal velocity and lateral velocity even gets worse compared to the ALDK. Meanwhile, the ALDK-FFRLS has the most degradation in modeling accuracy, with the maximum estimation error of yaw rate reaching 8.731 deg/s. Similarly, the 7 DoF-MF, which contains a maximum estimation error of 2.945 deg/s in yaw rate. As for the proposed ALDK-SWLS, the maximum estimation error of longitudinal velocity, lateral velocity, and yaw rate are 0.106 km/h, 0.011 km/h, and 0.377 deg/s, respectively, with comparison to the ground truth, highlighting a superior model fidelity.

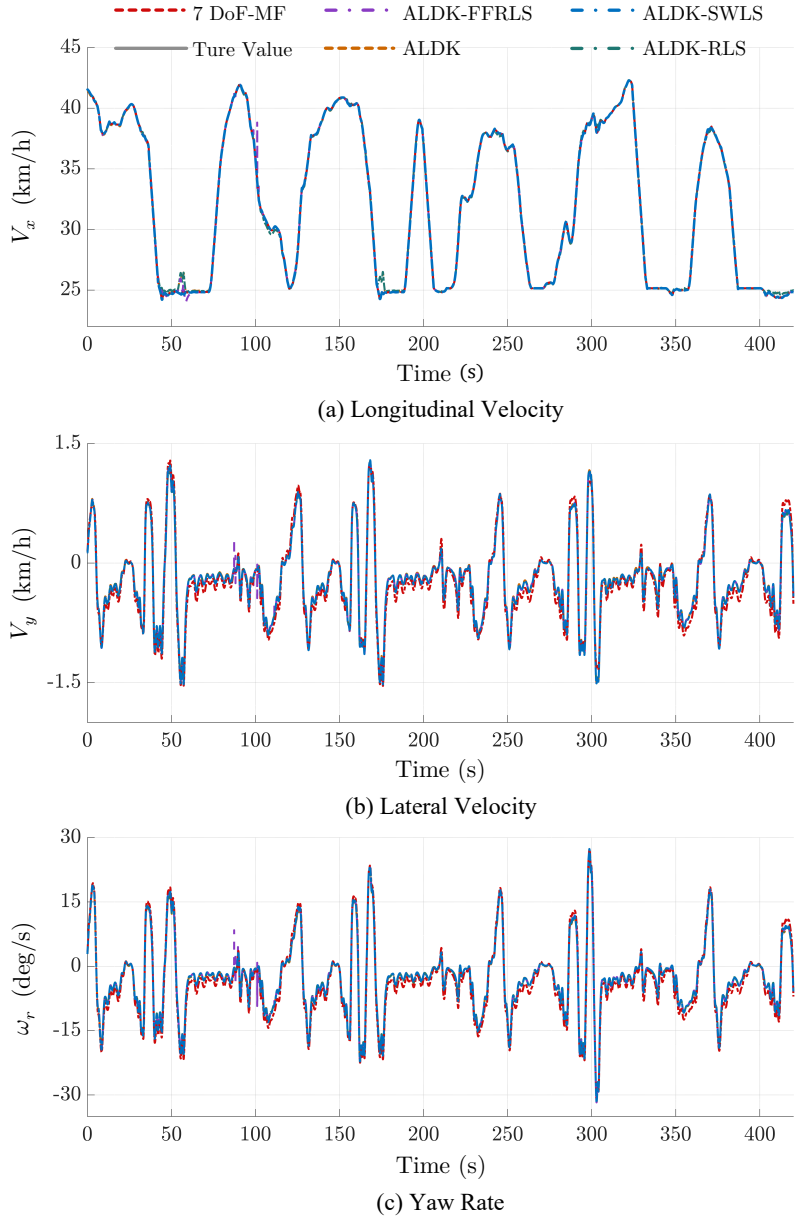

**Figure 5:** Simulation results of five approaches.

Because the ALDK-RLS and ALDK-FFRLS utilize all the historical data to update system matrices in the lifted space, the early-stage data information may not be consistent with the dynamics at the current time step, which causes estimation errors. Despite the higher modeling accuracy of the ALDK-RLS and ALDK-FFRLS methods, the noticeable oscillations during 50s to 100s in Figure 5 are not acceptable in real scenarios. In the 7 DoF-MF model, the inherent parameter uncertainties and unmodeled dynamics of the physics-based model-7 DoF-MF, such as rolling and aerodynamic resistances, lead to a loss of modeling accuracy. However, the ALDK-SWLS updates the system matrices based on the recent sequential data, thus capturing the current dynamics. This enables the ALDK-SWLS the outstanding online adaptability to the fast-varying dynamics of the IMDV and external environment. Moreover, the online recursive iteration of SWLS is computationally efficient.

**Table 4**
Statistic results of five approaches

| Estimation Error | $V_x$ | $V_y$ | $\omega_r$ |
|---|---|---|---|
| | Max/RMSE | Max/RMSE | Max/RMSE |
| 7 DoF-MF | 0.174/0.038 | 0.195/0.075 | 2.945/1.195 |
| ALDK | 1.839/0.220 | 0.053/0.014 | 1.157/0.282 |
| ALDK-RLS | 1.846/0.220 | 0.056/0.010 | 0.960/0.233 |
| ALDK-FFRLS | 4.827/0.108 | 0.371/0.007 | 8.731/0.154 |
| **ALDK-SWLS** | **0.106/0.011** | **0.011/0.001** | **0.377/0.025** |

**Table 5**
Statistical results of modeling errors with sprung mass change

| | Testing case 1: +160 kg | | | Testing case 2: -170 kg | | |
|---|---|---|---|---|---|---|
| | $V_x$ | $V_y$ | $\omega_r$ | $V_x$ | $V_y$ | $\omega_r$ |
| | Max/RMSE | Max/RMSE | Max/RMSE | Max/RMSE | Max/RMSE | Max/RMSE |
| 7 DoF-MF | 0.673/**0.045** | 0.270/0.077 | 3.904/1.263 | 0.673/0.051 | 0.337/0.074 | 0.854/0.178 |
| ALDK | 2.594/0.834 | 0.442/0.085 | 3.873/1.091 | 2.586/0.835 | 0.446/0.087 | 3.708/1.049 |
| ALDK-RLS | 2.556/0.818 | 0.433/0.076 | 4.037/0.912 | 2.557/0.819 | 0.436/0.080 | 3.866/0.891 |
| ALDK-FFRLS | 1.705/0.502 | 0.308/0.053 | 3.103/0.550 | 12.13/0.382 | 0.505/0.029 | 49.67/1.488 |
| **ALDK-SWLS** | **0.495**/0.088 | **0.067/0.006** | **1.384/0.122** | **0.247/0.043** | **0.097/0.021** | **1.282/0.231** |

### 4.3. Robustness Performance

Vehicle parameters may differ from those present in the collection. Therefore, we evaluate the robustness of the aforementioned five modeling methods in this section. The sprung mass and yaw inertial of the IMDV have been modified around the nominal value of the training dataset. In addition, the IMDV traverse along the *X-Y-Z edges* road to align with the Section 4.2.2.

#### 4.3.1. Varying Sprung Mass

We increase the sprung mass from 1648 kg to 1808 kg (+160 kg) and decrease it to 1478 kg (-170 kg), respectively. The increments of two cases are asymmetric to better reflect real-world events. The comparison and analysis are presented in Figure 6 and Table 5.

The ALDK and ALDK-RLS have similar performance in Figure 6(a) when the sprung mass increases. This indicates that the online adaption with the RLS does not improve the modeling accuracy than the unchanged one. As for the ALDK-FFRLS, the RMSE values for longitudinal velocity, lateral velocity, and yaw rate in the ALDK-FFRLS are 38%, 37%, and 49% lower in comparison to the ALDK. The proposed ALDK-SWLS stands out among these methods. The maximum estimation error of longitudinal velocity has decreased to 0.0668 km/h, lower than 0.4424 km/h of the ALDK and 0.2700 km/h of the 7 DoF-MF.

When the sprung mass decreases in Figure 6(b), the results and conclusions are similar. The ALDK-SWLS still has a higher modeling accuracy.

#### 4.3.2. Varying Yaw Inertia

Because of the different load distribution, the yaw inertia of the IMDV varies even with the same vehicle mass. As such, the yaw inertial has an increase of 142 kg · m² to 3800 kg · m² and a decrease of 158 kg · m² to 3500 kg · m² in this test. The performances of the five methods are validated in Figure 7 and Table 6.

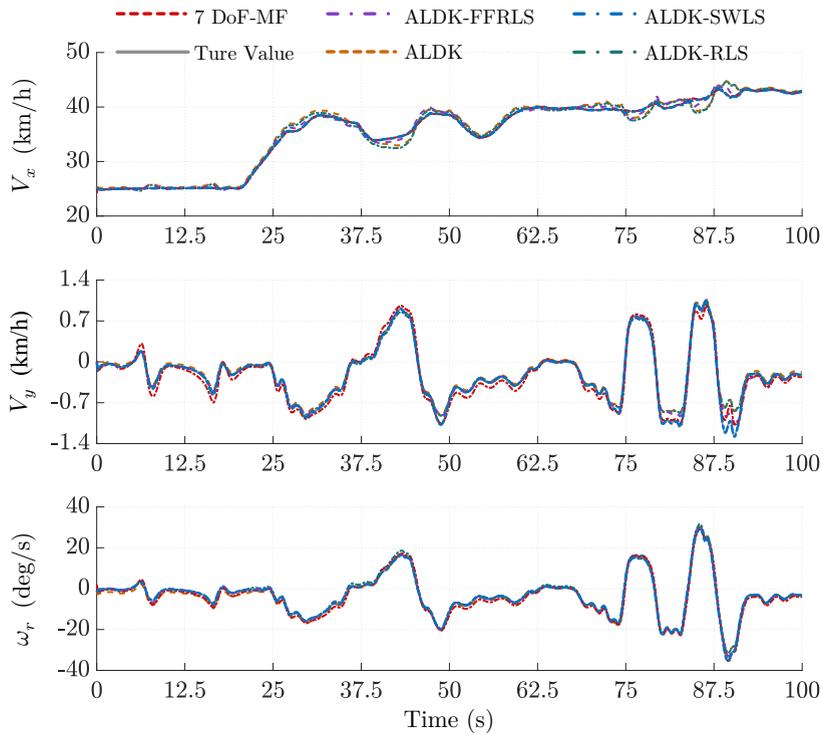

(a) Increase the spung mass by 160 kg

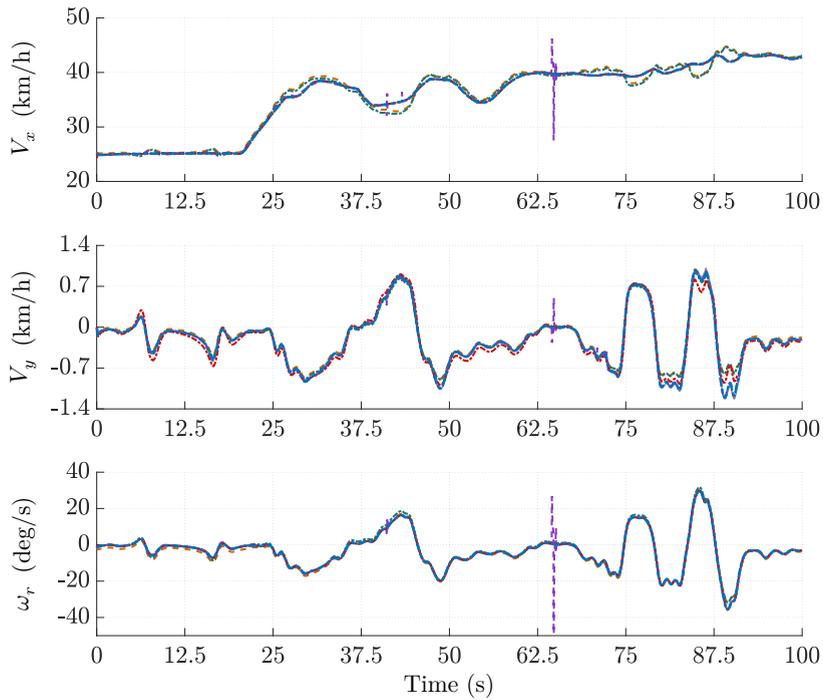

(b) Decrease the sprung mass by 170 kg

**Figure 6:** Testing results of varying sprung mass.

**Table 6**
Statistical results of modeling errors with yaw inertia change

|  | Testing case 1: $+142$ kg$\cdot$m$^2$ | | | Testing case 2: -158 kg$\cdot$m$^2$ | | |
| --- | --- | --- | --- | --- | --- | --- |
|  | $V_x$ | $V_y$ | $\omega_r$ | $V_x$ | $V_y$ | $\omega_r$ |
|  | Max/RMSE | Max/RMSE | Max/RMSE | Max/RMSE | Max/RMSE | Max/RMSE |
| 7 DoF-MF | 0.673/0.045 | 0.291/0.075 | 3.657/1.172 | 0.683/**0.032** | 0.287/0.072 | 2.963/1.085 |
| ALDK | 2.605/0.835 | 0.435/0.082 | 3.924/1.086 | 2.603/0.836 | 0.432/0.082 | 3.916/1.084 |
| fALDK-RLS | 2.573/0.822 | 0.427/0.074 | 4.078/0.916 | 2.570/0.822 | 0.423/0.074 | 4.072/0.914 |
| ALDK-FFRLS | 2.888/0.594 | 0.437/0.072 | 3.990/0.636 | 2.868/0.588 | 0.413/0.070 | 3.948/0.630 |
| **ALDK-SWLS** | **0.263**/**0.035** | **0.064**/**0.004** | **1.345**/**0.055** | **0.200**/0.038 | **0.058**/**0.005** | **1.217**/**0.070** |

When the yaw inertia increases, the ALDK-FFLS still has an obvious variation, but the ALDK-SWLS has a significant improvement than other approaches in Figure 7(a). Compared to the 7 DoF-MF, the maximum estimation error of the ALDK-SWLS for longitudinal velocity decreases from 0.673 km/h to 0.263 km/h while exhibiting a 63% reduction for yaw rate. Meanwhile, the maximum values of the ALDK-SWLS all arise at the initial moment because it degrades to the RLS. In terms of the RMSE, the value are 0.835 km/h, 0.082 km/h, and 1.086 deg/s of the ALDK, while there are 0.035 km/h, 0.004 km/h, and 0.055 deg/s of the ALDK-SWLS. Similarly, the results are the same for the decreasing case in Figure 7(b) and Table 6.

In fact, the accuracy of the physics-based model (the 7 DoF-MF) is subject to the precision of the vehicle parameters . The discrepancy between these parameters used in the physics-based model and their corresponding real-world values has an impact on the model fidelity. Based on the analysis in Sections 4.3.1 and Section 4.3.2, it is proven that the ALDK-SWLS has satisfactory robustness to the variations of sprung mass and yaw inertia.

### 4.4. Generalization Performance

Actually, the real-world roads environment is more complex than the *X-Y-Z edges* road. To further validate the adaptability of the proposed ALDK-SWLS, two typical roads – mountain road and slalom test road, with features different from the *X-Y-Z edges* road, are configured in CarSim to conduct the generalization test in this section.

#### 4.4.1. Mountain Road

The mountain road has large elevational drops and extreme cornering conditions. The IMDV not only generates longitudinal and lateral motions, but also has response a vertical response. The initial longitudinal velocity is 20 km/h, and the road adhesion coefficient is 0.85.

The results in Figure 8 reveal that the ALDK, ALDK-RLS, and ALDK-FFRLS show significant errors from the actual values, especially in the longitudinal velocity, where the maximum estimation errors in Table 7 are 4.813 km/h, 5.831 km/h and 2.749 km/h, respectively. The performance of the three models for lateral velocity and yaw rate predictions is also unsatisfactory. Meanwhile, the RMSE value for yaw rate of 7 DoF-MF is 1.150 deg/s. This is because the actual vehicle behaviors are far from the assumptions within the 7 DoF-MF model. However, due to the receding horizon of SWLS, the ALDK-SWLS captures the instant vehicle dynamics well. The maximum estimation errors of longitudinal velocity and yaw rate are 0.112 km/h and 0.497 deg/s, and the corresponding RMSE values are 0.001 km/h and 0.007 rad/s, respectively.

#### 4.4.2. Slalom Test Road

In this case, the maximum velocity is 120 km/h with an initial value is 20 km/h, and the road adhesion coefficient is 0.6. The road surface is flat, but the steering wheel angle is set as a sinusoidal input with periodical time of 100 s to induce high-frequency responses of the IMDV.

Due to the obvious discrepancy between the training features and this testing scenario, Figure 9 and Table 8 present results that are remarkable. Similar to Section 4.4.1, the maximum estimation errors of the longitudinal velocity for the ALDK, ALDK-RLS, and ALDK-FFRLS are 11.120 km/h, 9.072 km/h, and 8.026 km/h, respectively. During the time 15-50s, the steering maneuver causes rapid responses in lateral velocity and yaw rate, and therefore, it induces great disturbance to the longitudinal velocity estimation. In contrast, the vehicle states change slowly across 50-100s. As a result, there are no significant deviations between the lateral velocity and yaw rate estimations of the four data-driven methods and the actual values. The 7 DoF-MF model has a more accurate longitudinal velocity estimation than the ALDK-SWLS, while its estimations on lateral motions, especially the yaw rate, get worse with the maximum value of 2.195 deg/s. Nevertheless, the ALDK-SWLS certifies a performant and reliable modeling method.

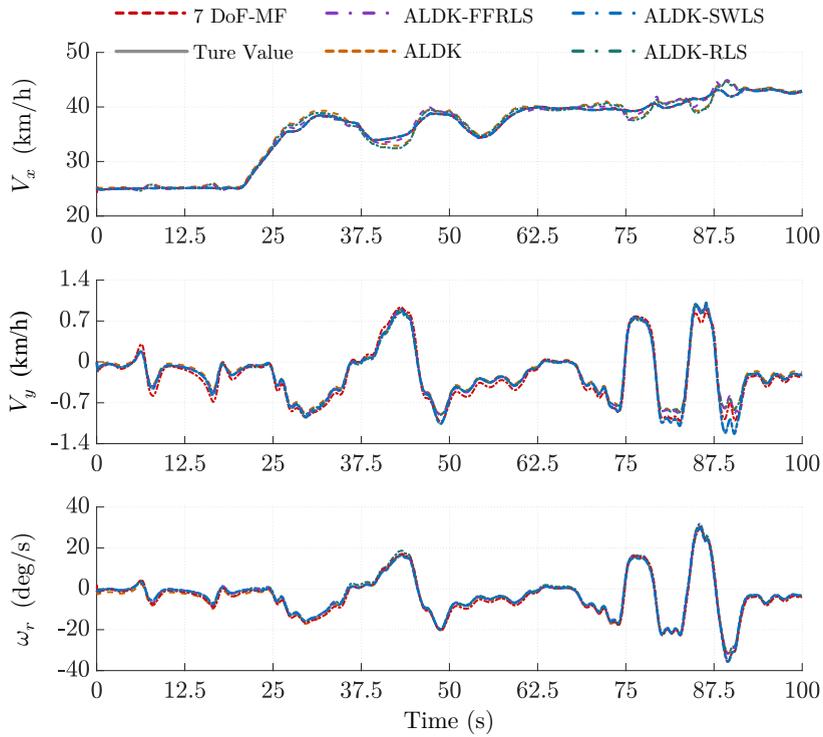

(a) Increase the yaw inertia by 142 kg·m²

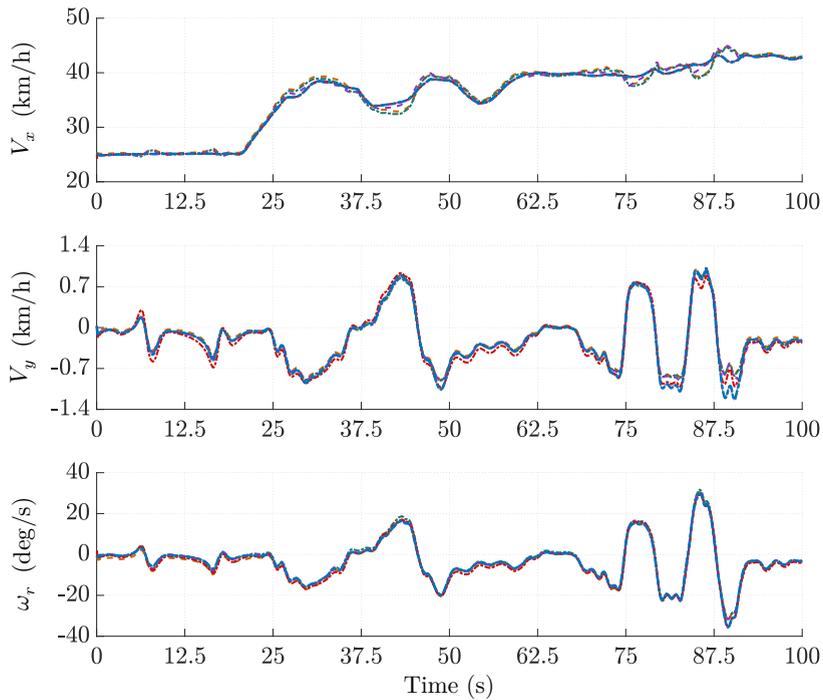

(b) Decrease the yaw inertia by 158 kg·m²

**Figure 7:** Testing results of varying vehicle yaw inertia.

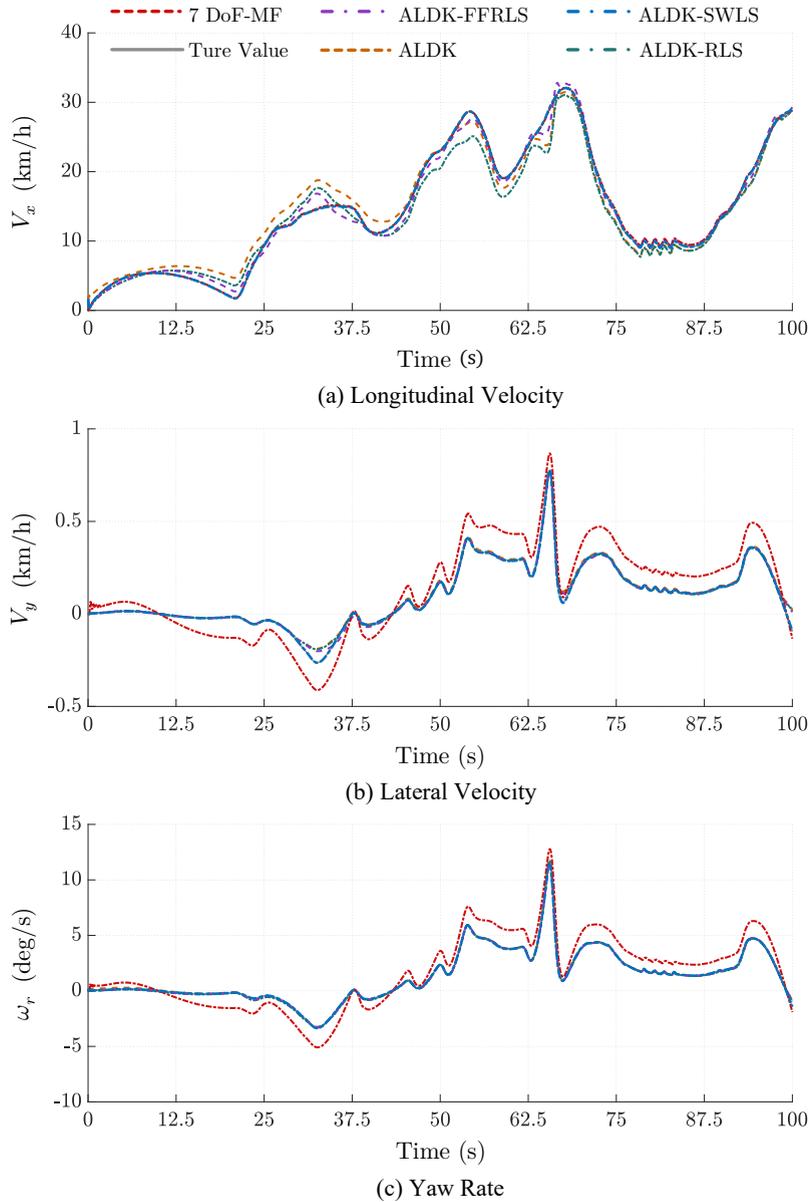

**Figure 8:** Simulation results of mountain road.

In general, based on the discussions above, the ALDK-SWLS approach has the best overall performance both in the fast and steady dynamics. The acceleration loss-informed deep neural network refines the accuracy of Koopman operator approximation and renders it with inherent generalization. The application of SWLS enforces the deep Koopman operator capability to cope with changes in vehicle parameters, road conditions, and rapid maneuvers. In general, based on the discussions above, the ALDK-SWLS approach has the best overall performance in both the fast and steady dynamics. The acceleration loss-informed deep neural network refines the accuracy of Koopman operator approximation and renders it with inherent generalization. The application of SWLS renders the deep Koopman operator to cope with changes in vehicle parameters, road conditions, and rapid maneuvers.

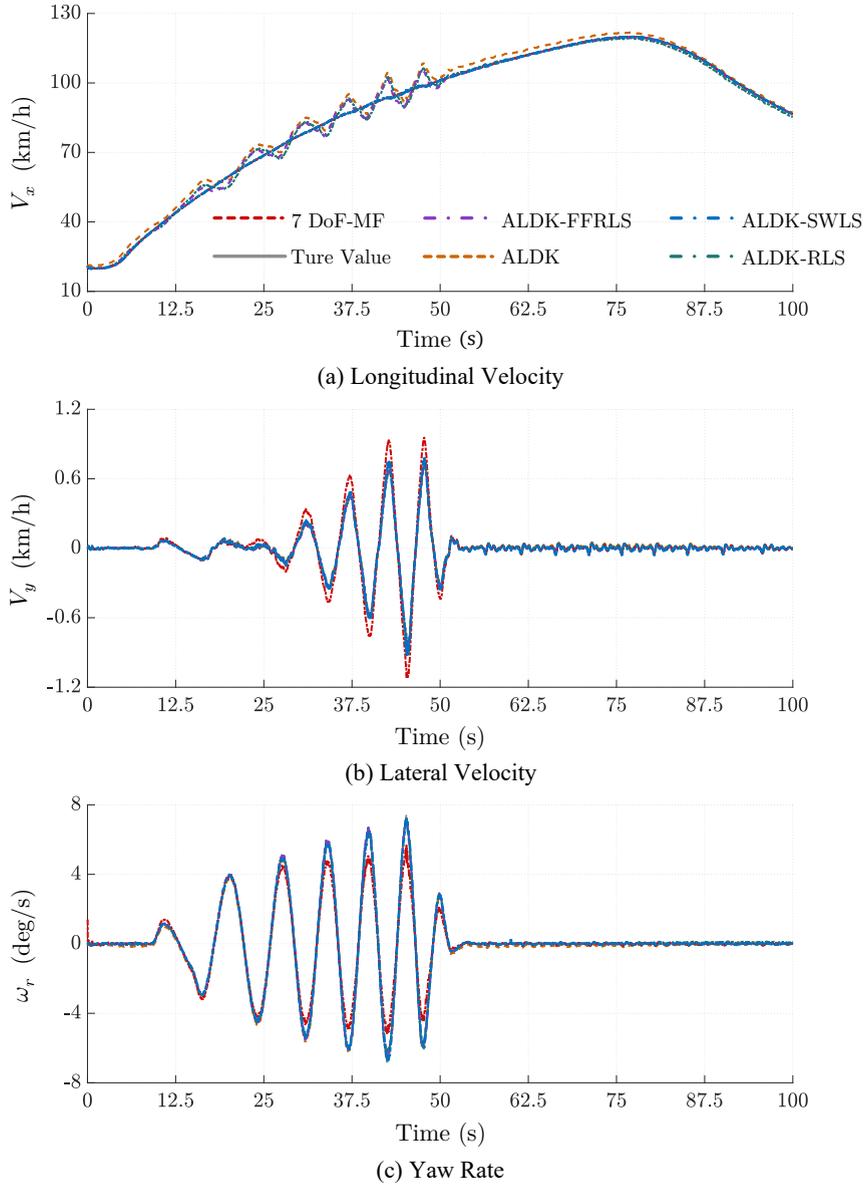

**Figure 9:** Simulation results of Slalom test road.

## 5. CONCLUSION

This study develops the physics-informed adaptive deep Koopman operator as a data-driven modeling tool for autonomous vehicle dynamics. The DNNs are employed to approximate basis functions of Koopman operator and to feature system and input matrices in the lifted space. The acceleration loss between network output and sensor measurements is included as the physics-informed term in the total training loss. Besides, the SWLS method is applied to recursively update the system matrix and input matrix in the lifted space online. Further, the validation and comparison are conducted on the CarSim/Simulink co-simulation platform with other physics-based and date-driven methods under various vehicle parameters, road conditions, and rapid maneuvers.

Table 7
Statistic results of mountain road

| Estimation Error | $V_x$ | $V_y$ | $\omega_r$ |
|---|---|---|---|
| | Max/RMSE | Max/RMSE | Max/RMSE |
| 7 DoF-MF | **0.345**/0.146 | 0.156/0.098 | 2.323/1.150 |
| ALDK | 4.813/1.622 | 0.118/0.018 | 0.507/0.083 |
| ALDK-RLS | 5.831/1.622 | 0.112/0.016 | 0.497/0.081 |
| ALDK-FFRLS | 2.749/0.776 | 0.105/0.015 | 0.495/0.063 |
| **ALDK-SWLS** | 1.678/**0.044** | **0.006/0.001** | **0.181/0.007** |

Table 8
Statistic results of Slalom test road

| Estimation Error | $V_x$ | $V_y$ | $\omega_r$ |
|---|---|---|---|
| | Max/RMSE | Max/RMSE | Max/RMSE |
| 7 DoF-MF | **0.562/0.027** | 0.242/0.052 | 2.195/0.448 |
| ALDK | 11.12/2.994 | 0.143/0.017 | 0.540/0.145 |
| ALDK-RLS | 9.072/2.271 | 0.139/0.014 | 0.592/0.096 |
| ALDK-FFRLS | 8.026/2.035 | 0.129/0.012 | 0.495/0.083 |
| **ALDK-SWLS** | 1.073/0.173 | **0.092/0.009** | **0.350/0.080** |

Firstly, compared to the traditional deep Koopman operator model (DK), the acceleration loss-informed deep Koopman operator (ALDK) extracts the interpretable information of the dataset and refines the accuracy of Koopman operator approximation. Secondly, incorporating the ALDK with the SWLS, the proposed ALDK-SWLS can update the system matrices from the recent sequential data and thus capture the current dynamics, which enables it an outstanding real-time adaptability to the fast-varying dynamics than the 7 DoF-MF (physics-based model), ALDK, ALDK-RLS, and ALDK-FFLS. As such, the ALDK-SWLS exhibits overall satisfactory robustness and generalization performances in autonomous vehicle dynamics estimation and is a potential data-driven modeling tool.

The main work of this paper focuses on the construction of the vehicle dynamics model, and future works will explore the implementation of vehicle state observers and stability controllers using the proposed model.

## Funding

This work was supported by the graduate research and innovation foundation of Chongqing, China (Grant No. CYS23045).